\documentclass[twocolumn,nofootinbib,showkeys,aps]{revtex4}
\usepackage{latexsym}
\usepackage{graphicx}  
\usepackage{dcolumn}   
\usepackage{bm}        
\usepackage{color}
\usepackage{setspace}

\newcommand{\gb}{{\rm gb}}
\newcommand{\BC}{{\rm B_{\mbox{\tiny \rm C}}4}}

\begin{document}

\title{
Combined volumetric, energetic and microstructural defect analysis
of ECAP-processed nickel\\
}

\author{Gerrit Reglitz$^1$} 
\author{Bernd Oberdorfer$^{2,3}$}
\author{Nina Fleischmann$^2$}
\author{Jaromir Anatol Kotzurek$^2$}
\author{Sergiy V. Divinski$^{1,*}$}
\author{Wolfgang Sprengel$^{2,**}$}
\author{Gerhard Wilde$^1$}
\author{Roland W\"{u}rschum$^2$}
\affiliation{
$^1$Institute of Materials Physics; University of M\"{u}nster; D-48149 M\"{u}nster, Germany \\
$^2$Institute of Materials Physics; Graz University of Technology; A-8010 Graz, Austria \\
$^3$now at: Austrian Foundry Research Institute; A-8700 Leoben, Austria \\
Corresponding authors: $^*~$divin@wwu.de \\
$~$\hspace{4.5cm} $^{**}$w.sprengel@tugraz.at
            }

\date{\today}

\begin{abstract}
\noindent Difference dilatometry and differential scanning calorimetry (DSC) are used to investigate defect annealing in ultrafine grained nickel processed by equal channel angular pressing (ECAP) at various temperatures. Different defect types and processes such as  vacancies, dislocations, grain boundaries and grain-boundary relaxation can be detected. They can be distinguished due to their distinct kinetics as revealed by the release of excess volume and excess heat during linear heating. The data are quantified in combination with a detailed characterization of the microstructure. Values for the absolute vacancy concentration, the dislocation density, the grain boundary expansion and the excess of grain boundary expansion in ECAP-processed nickel are derived.
\end{abstract}



\maketitle

\section{Introduction}
\noindent Ultrafine-grained (UFG) materials with sub-micron grain size and attractive physical properties can be produced by techniques of severe plastic deformation (SPD) such as Equal Channel Angular Pressing (ECAP) or High Pressure Torsion (HPT) \cite{book}. The specific preparation technique, e.g., the ECAP preparation routes, as well as post-deformation annealing treatment significantly affect the physical properties and processes of these UFG materials. For example it was shown that diffusion and especially the temperature dependence of grain boundary diffusion in ECAP-processed Ni are quite intricate \cite{Divinski2011, Divinski2015, Divinski2014}. Those kind of influences are most probably due to a complex defect structure and/or specific defect interaction and defect annealing behavior. 

In the case of UFG-Ni produced by HPT, the annealing of defects was investigated by Setman et al. using calorimetry, resistivity measurements and X-ray Bragg profile analysis \cite{zehet}. That work was performed on high-purity, $99.99$~wt.\%, Ni and recrystallization considerably affected the heat release from dislocation reactions \cite{zehet, zehet2}. Furthermore, UFG high-purity Ni produced by HPT was analyzed by difference-dilatometry \cite{Sprengel2012} to derive values for the grain boundary expansion \cite{Steyskal2012} or to study defect annealing \cite{OberdorferPRL} or the recrystallization kinetics \cite{Oberdorfer2011}. Recently volumetric changes at isochronal annealing of HPT-processed Cu were reported \cite{Cu}.
 
The objective of the present investigation is to combine volumetric (dilatometry) and energetic (calorimetry) studies with microstructure studies to characterize the different defect types and to study the temperature regimes of their annealing in detail. Furthermore, the measurements were extended to specimens with post-deformation annealing, i.e., pre-annealing with respect to the measurement, as well as to specimens that had been deformed at elevated temperatures (see Tab.~\ref{table}). This unique combination of different analysis methods and different specimen conditions allows for a comprehensive estimation and determination of values for basic defect parameters such as: the absolute vacancy concentration, the dislocation density, the grain boundary expansion and the excess of grain boundary expansion, here especially applied to the case of ECAP deformed $99.6$~wt\% Ni. The data are critically discussed with respect to data on kinetic properties of interfaces and defect mobilities along short-circuit paths in the same Ni material determined previously \cite{Divinski2011}. The microstructure characterization is shown to be essential for rigorous analysis of the volumetric effects. 

\begin{table*}[ht]%
\caption{\footnotesize Summary of specimens investigated by dilatometry. Measurements were performed in the temperature range from $293$~K to $773$~K.}\label{table}
\medskip
\begin{tabular}{ccccc}
  \parbox{3cm}{Code}     
& \parbox{3cm}{deformation temperature \\ (K)} 
& \parbox{3cm}{pre-annealing temperature \\ (K)} 
& \parbox{3cm}{pre-annealing time \\ (h)}
& \parbox{3cm}{dilatometry heating rate (K/min)}\\
\hline \hline
&&&&\\
P3 & 293 & --  & --&1.5, 3.0, 6.0 \\
&&&&\\
P4 & 293 & 373 &48& 3.0 \\
P5 & 293 & 480 &17& 3.0 \\ 
&&&&\\
P6 & 473 & -- & -- &3.0 \\
P7 & 673 & -- & -- &3.0 \\
\end{tabular}
\end{table*}

\section{Experimental Procedure}

\noindent Ultrafine-grained nickel specimens were prepared by ECAP techniques from nickel of nominally $99.6$\% purity in the research laboratory of Y. Estrin (Monash University, Clayton, Australia). The chemical analysis was performed by the supplier and the results are shown in Table \ref{tab:impurity}. The usage of technically pure Ni in the present work allowed to separate qualitatively the stages of vacancy annihilation, dislocation evolution and recrystallization due to additional pinning forces which were exerted by impurities, see below. Moreover, as it was stated above, this is exactly the same material which kinetic properties and microstructure stability were extensively examined in our previous works \cite{Divinski2011, Divinski2015, Divinski2014}.

The $\BC$ route was used where four ECAP passes are employed and the specimens are rotated by $90^\circ$ in the same direction after each pass. Rods of $10$~mm in diameter were deformed at three different temperatures, $293$~K, $473$~K, and $673$~K. As a result, specimens with UFG microstructures were produced.

\begin{table}[ht]%
\caption{\footnotesize Impurity concentration (in wt.ppm) in the Ni material under investigation.} \label{tab:impurity}
\begin{tabular}{c|c|c|c|c|c|c|c|c|c|c}
 \parbox{0.6cm}{Al}  &  \parbox{0.6cm}{Co}  & \parbox{0.6cm}{Cr} & \parbox{0.6cm}{Cu} &  \parbox{0.6cm}{Fe}  &  \parbox{0.6cm}{Mn}  & \parbox{0.6cm}{Mg} & \parbox{0.6cm}{P}  &  \parbox{0.6cm}{Si}  &  \parbox{0.6cm}{Ti}  & \parbox{0.6cm}{Zn} \\ \hline
$70$ & $100$ & $15$ & $380$& $350$&$1100$& $20$ & $30$& $170$& $30$ & $4$\\ 

\end{tabular}
\end{table}

Furthermore, prior to the dilatometric measurements, two UFG specimens were pre-annealed at $373$ and $480$~K for $48$ and $17$~hours, respectively. As will be seen below these treatments produce specific defect configurations which allow to quantify different stages of microstructure evolution. The specifications of all samples used for the dilatometric measurements are given in Table~\ref{table}.

\subsection{Dilatometry measurements}

\begin{figure}[ht]
\begin{center}
\includegraphics[width=8cm]{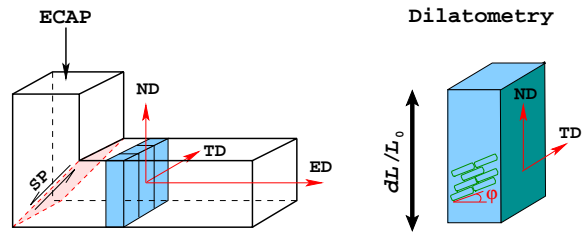}
\end{center}
\caption{\footnotesize Sketch and definition of the orientations of the ECAP process (left panel) and of the dilatometric measurement (right panel): ED is the extrusion direction, ND the normal direction, TD the transverse direction, $dL/L{_0}$ the dilatometric measurement direction, $\phi$ the inclination angle of the elongated grains with respect to the ED-TD plane as determined by microstructure study, and SP is the position of the ideal shear plane.} \label{fig:sketch}
\end{figure}

\noindent For the dilatometric measurements prism-shaped samples ($3 \times 3 \times 8$~mm$^3$) were cut with the long axis being parallel to the measurement direction and oriented perpendicular to the ECAP processing direction, i.e., the extrusion direction, ED, and parallel to the normal direction, ND, direction, see Fig.~\ref{fig:sketch}.

The experiments were performed with a high-precision, vertical double-dilatometer (Linseis, L75VD500 LT) allowing a simultaneous  measurement of two specimens. In the present case, the second specimen which served as reference was a coarse-grained specimen (with a mean grain size of about $100\mu$m). The reference specimen was prepared from the same Ni material but annealed at $1273$~K and slowly cooled over several days. Note that exactly this state of Ni material was used as initial for present ECAP processing. Thus, it can safely be stated that the only difference between the reference and UFG specimens corresponds to the defects introduced by ECAP deformation. The final experimental data, as represented in a dilatometric length change curve, $\Delta L / L_0$, is the difference between the specimen signal and the reference signal (difference dilatometry). In this way, the length change effects due to linear thermal expansion caused by the lattice anharmonicity are canceled out and the defect-related effect is measured exclusively. It should be noted that it has also been experimentally verified~\cite{Kotzurek2013} that within experimental uncertainties the ultrafine-grained sample with the complex defect structure has the same reversible linear thermal expansion coefficient as the coarse-grained, well-annealed reference sample. Furthermore, via the relation $3 \cdot \Delta L / L_0$ = $\Delta V / V_0$ the length change is directly related to the defect volume if an isotropic distribution and annealing of defects can be assumed \cite{Sprengel2012}.

By calculating the derivative of the length change curve with respect to the temperature the temperatures with maximum defect-release-rates can be determined from the minima of the $d(\Delta L/L_0)/dT$ curve. This is valid as by employing constant linear heating rates, $dT/dt$, the temperature, $T$, is direct proportional to the time, $t$.

The dilatometer operates mechanically with two independent quartz push-rods and allows absolute length change measurements $\Delta L$ of the order of $30$~nm. For specimens of $8$~mm in length, this gives access to absolute changes in free volume fractions down to about $\Delta V / V_0 = 1 \cdot 10^{-5}$. Measurement of the sample/reference pair was performed using a constant, high-purity argon gas (5N) flow inside the sample chamber. For the study of the defect annealing kinetics usually measurements with a series of different linear heating rates are performed. In this case the heating rates of $1.5$, $3$, and $6$~K/min were used for the specimens denoted P3. All other specimens were measured with a heating rate of $3$~K/min. For the final temperature readings a correction is applied which takes a varying temperature delay for different heating rates into account and which is caused by the specific set-up of the dilatometer. For details of this correction see Ref.~\cite{Oberdorfer2010}.

\begin{figure*}[ht]
\begin{center}
a) \includegraphics[width=5cm]{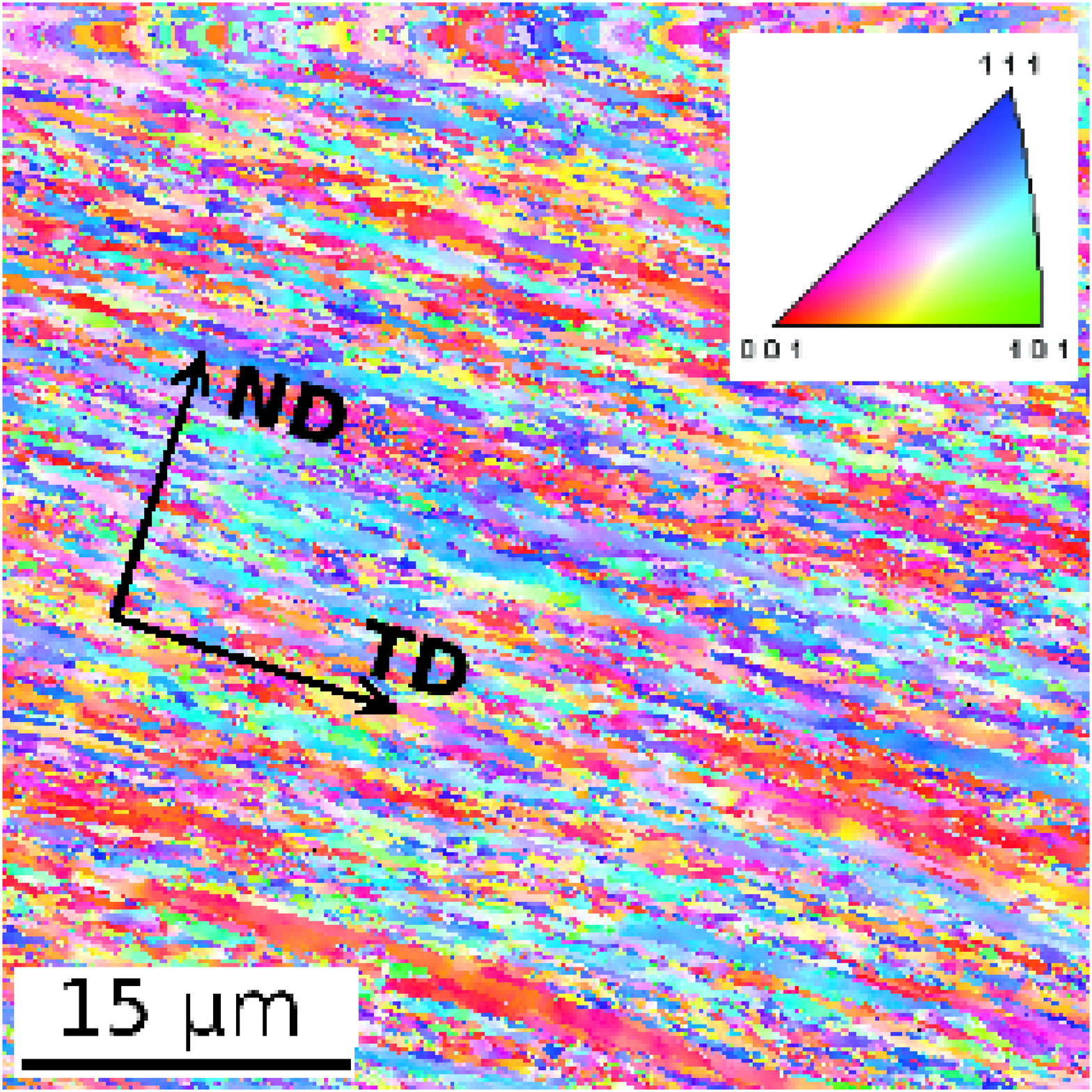}
\hspace{0.3cm}
b) \includegraphics[width=5cm]{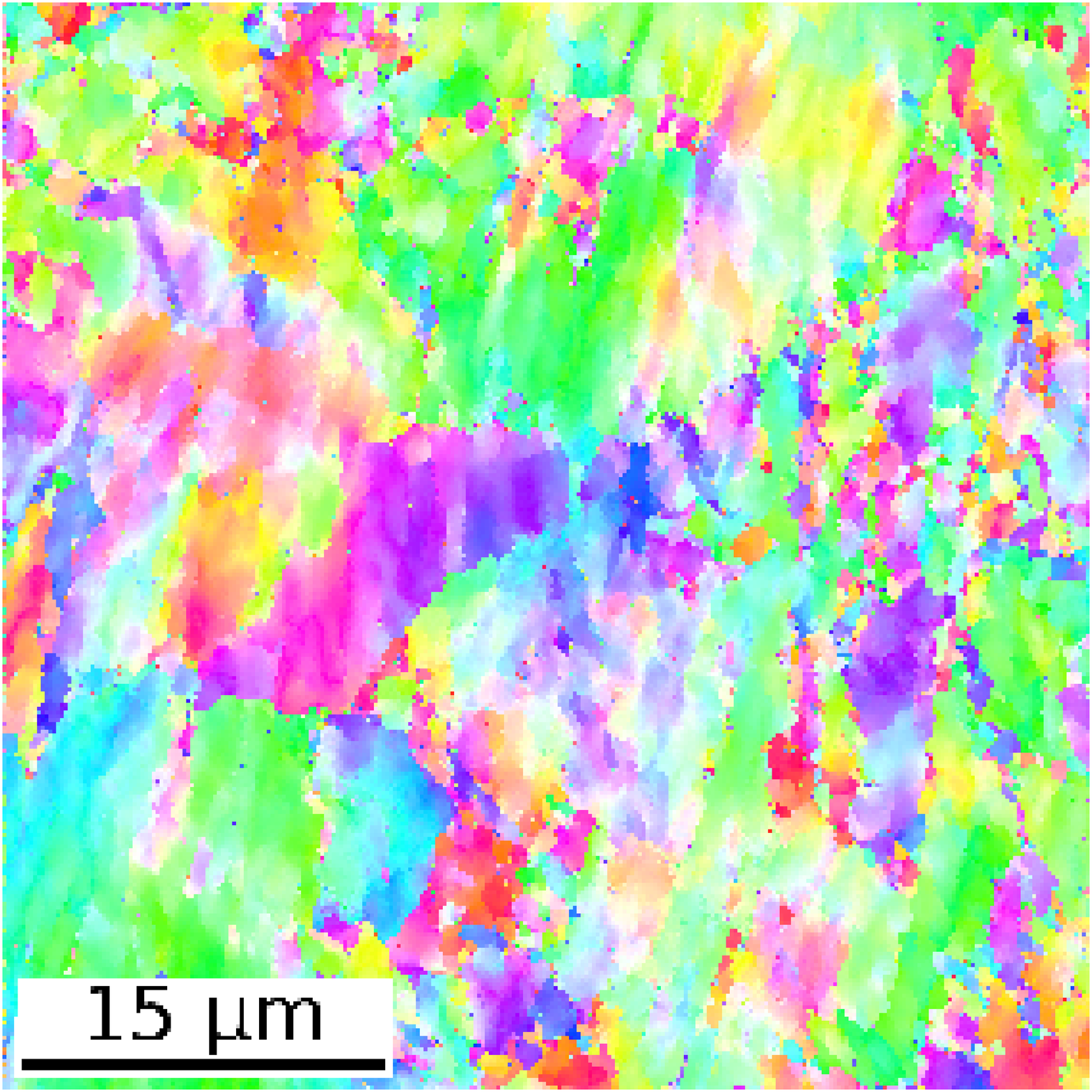}
\hspace{0.3cm}
c) \includegraphics[width=5cm]{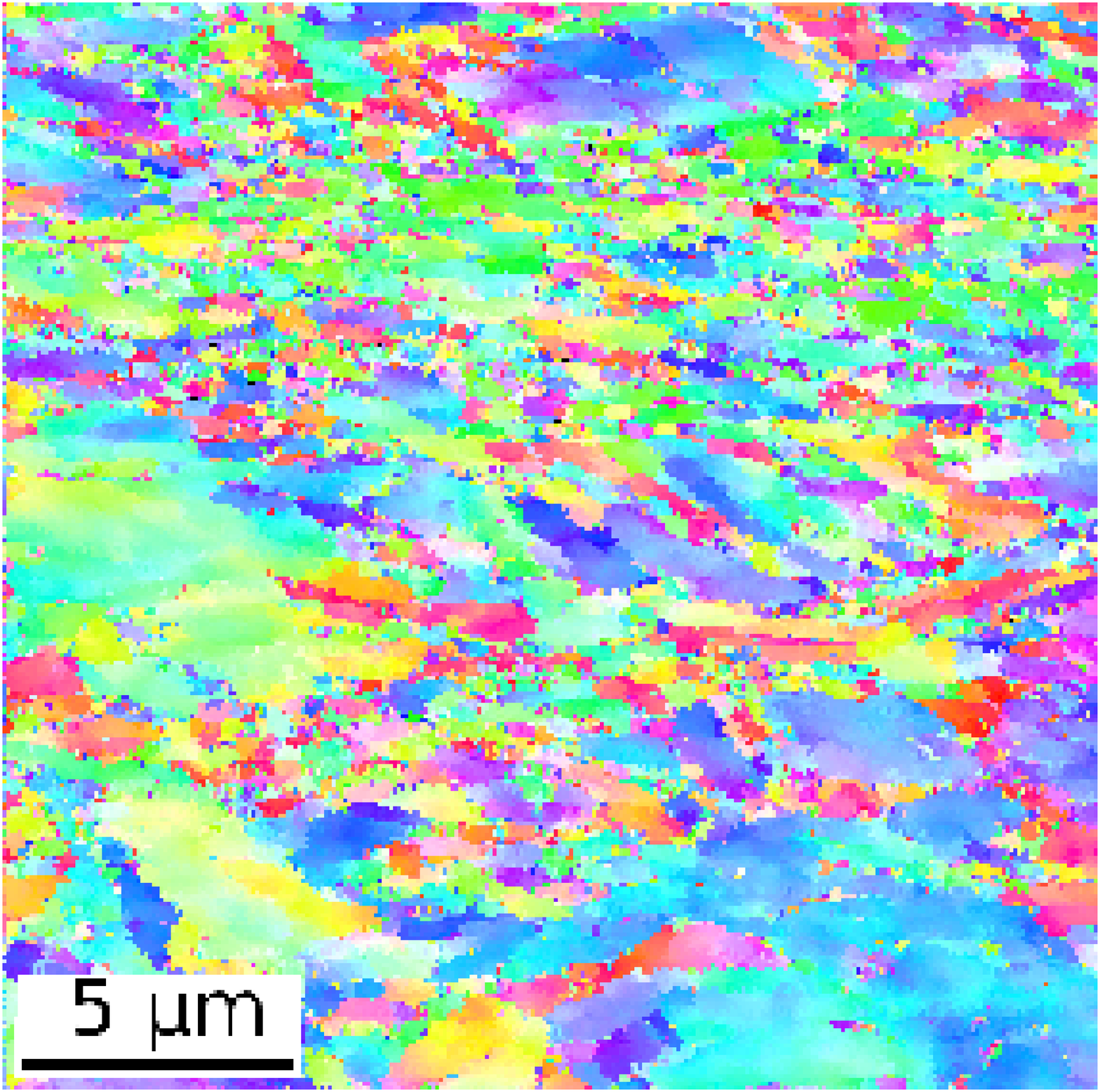}
\end{center}
\vspace*{-3mm}
\caption{Orientation imaging microscopy of the initial microstructure of Ni deformed by $\BC$-type ECAP processing at $293$~K (a), $473$~K (b), and at $673$~K (c). The grain color corresponds to inverse pole figure coloring as indicated in the insert of (a). By ND and TD the normal and the transverse directions of the ECAP processing are denoted, see Fig.~\ref{fig:sketch}.} 
\label{fig:micro}
\end{figure*}

In all measurement runs the specimens were first cooled down to $273$~K from where the measurements start. This procedure assures that already slightly above room temperature, the dilatometer is in a thermally stable condition with the preset, constant heating rate, allowing reliable analysis of defect annealing at temperatures as low as $323$~K. 

\subsection{Calorimetric measurements}

\noindent Calorimetric measurements were performed by differential scanning calorimetry (DSC, Perkin Elmer, USA) under Ar atmosphere. The samples were disc shaped with a mass of about $160$~mg. The DSC scans were recorded in the temperature interval from $323$~K to $723$~K for different heating rates ranging from $5$~K/min to $80$~K/min. For each sample, three identical DSC runs were performed. The second run was used as a baseline, which was then subtracted from the first run to obtain only the irreversible part of the heat flow signal. Within experimental uncertainties, the second and the third scans yielded nearly identical signals. Thus, the differential heat flux data correspond to irreversible processes which are completed during the first run. A standard temperature correction was applied for each heating rate. Similarly to the dilatometric experiments, the samples were first cooled down to $273$~K from where the measurements started. 

Isothermal measurements of the heat release were performed with a Thermal Activity Monitor (TAM) III multi-channel isothermal micro calorimeter (TA Instruments). In such a case, the sample is first equilibrated slightly below the measurement temperature for about $15$~min and then shifted to the measurement position. After about $30$ to $45$~min all transient processes are accomplished and the heat release is then recorded.

\subsection{Microstructure characterization}

\noindent The microstructure was characterized by scanning electron microscopy (SEM) using a FEI Nova NanoSEM 230 device. Electron back-scatter diffraction (EBSD) was used to identify the grains and to determine the grain orientation. The typical lateral scan step was $60$~nm and several areas of at least $45 \times 45~\mu$m$^2$ were inspected on each specimen to provide the sample-averaged microstructure data.

The initial microstructure of the UFG Ni specimens is shown in Fig.~\ref{fig:micro} for all three deformation conditions. Orientation imaging microscopy is performed for sections perpendicular to the extrusion direction. The microstructure is typical for SPD deformed Ni, see e.g. \cite{Pippan1, Pippan2}.
Elongated grains are formed after ECAP deformation at room temperature (Fig.~\ref{fig:micro}a), while more equiaxial grains are produced by warm deformation. The significant impact of recovery processes during warm deformation at $473$~K and $673$~K is directly visible in Fig.~\ref{fig:micro}b, c. The room temperature (293~K) deformation induces numerous shear (deformation) bands with boundaries mainly oriented parallel to the transverse direction (see Fig.~\ref{fig:micro}a). Warm deformation, however, results in a bimodal microstructure with a UFG fraction (the grain size of about $350$ to $500$~nm) and larger recrystallized grains of $3$ to $5~\mu$m in diameter with a more isotropic grain shape. Furthermore, a significantly more heterogeneous microstructure is formed after warm deformation (especially at $473$~K) that indicates recrystallization and abnormal grain growth processes already during straining and / or during the post-deformation cooling phase. Correspondingly, alongside with a UFG microstructure component, relatively large grains, several micrometers in diameter, could be found, Fig.~\ref{fig:micro}b, c.

\section{Results}

In the present work two different techniques are basically applied to analyze the defect evolution under linear heating conditions. A fundamental problem appears when the results of these techniques are compared, if different heating rates are used that is almost unavoidable in view of different sensitivities of the corresponding apparatus. In this case, characteristic processes might be not completed during dilatometry and DSC measurements at the same temperatures and thus they will contribute differently to other processes occurring at higher temperatures. Therefore, in the present paper the defect analysis is primary based on the dilatometry data. 

The above remarks can be applied to comparison of as-deformed and pre-annealed (at a constant temperature for a given time) samples, too. Therefore, when such a comparison will be done in the present manuscript, it will be concentrated on the processes that occurred \emph{below} the pre-annealing temperature.

\subsection{Difference dilatometry}

\begin{figure}[ht]
\vspace{1.5cm}

\includegraphics[width=7cm]{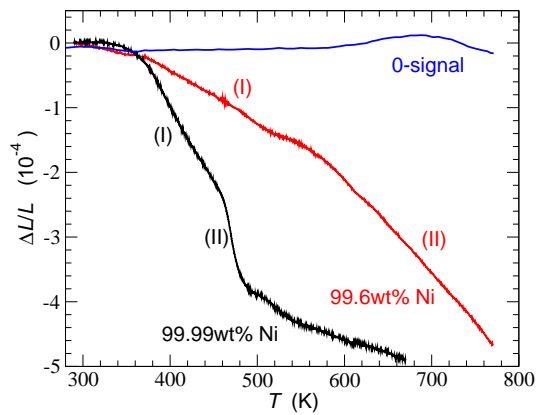}
\caption{\footnotesize Relative length change, $\Delta L / L_0$, upon heating with a linear rate of $3$~K/min of an ECAP-deformed Ni sample P3 (99.6 wt\%) (red line) in comparison to HPT-deformed Ni (99.99 wt\%) (black line \cite{Sprengel2012}). Two distinct temperature ranges of free volume annealing are marked by (I) and (II); for details see text and further analysis. The blue line represents the 'zero-effect' signal measured between two reference samples.} \label{fig:compHPT}
\end{figure}

\noindent Figure~\ref{fig:compHPT} shows the length change curves, $\Delta L / L_0$, upon heating with a linear heating rate of $3$~K/min of an ECAP-Ni sample deformed at 293~K~(P3). The data are compared to the length changes in an HPT-deformed high-purity Ni (99.99 wt\%) sample studied before \cite{Sprengel2012} (black). The blue curve in Fig.~\ref{fig:compHPT} represents the zero-effect measurements in which two reference samples were measured against each other. This curve substantiates that the signals measured for UFG samples exceed the zero-signal by, at least, an order of magnitude.

The length changes due to the defect annealing in the ECAP sample are clearly seen to be significantly shifted to higher temperatures compared to the $99.99$~wt\% purity HPT-deformed Ni sample. For HPT-processed Ni, at least two distinct annealing stages were observed independent on the applied total strain, the stages are denoted (I) and (II) \cite{Sprengel2012}. The steep change in stage (II) was explained by grain growth, whereas stage (I) was attributed to relaxation of grain boundaries, annealing of vacancies and dislocations \cite{OberdorferPRL, Oberdorfer2011}. It seems that both stages occur in ECAP-deformed low-purity Ni, too, however, in a much wider temperature range and shifted to higher temperatures. The retardation of recrystallization and grain growth processes in $99.6$ wt.\% Ni might be explained by impurity drag effects, see, e.g., Ref.~\cite{Humphreys}, which give rise to an extended stage (I) with important sub-stages as explained below.

\begin{figure}[ht]
\includegraphics[width=7cm]{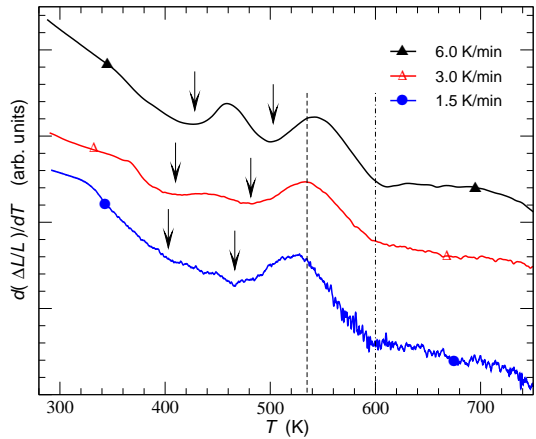}\\
\caption{\footnotesize Temperature derivatives of the length change curves, $d(\Delta L/L_0)/dT$, measured on sample P3 with different heating rates. Below about $535$~K (the vertical dashed line) two different annealing processes can be distinguished in all three curves and which are indicated by arrows. For temperatures above $535$~K a process similar for all three heating rates starts, which reveals a two-stage character with respect to the rate of the length changes -- below and above about $600$~K (the vertical dashed-dotted line) -- and it is attributed to grain growth. Here and below the temperature derivative curves are shifted one respect to another for a better visualization and the spacing between the major tics on ordinate axis corresponds to $10^{-6}$~K$^{-1}$.} \label{fig:deriv}
\end{figure}

The dilatometry measurements on ECAP-Ni were repeated with the heating rates of $1.5$ and $6$~K/min. The total release of defect volume is largest at the lowest heating rate and higher heating rates shift the volume release to higher temperatures as is expected for thermally activated processes. The temperature derivatives of the measured relative length changes, $d(\Delta L/L_0)/dT$, are plotted in Fig.~\ref{fig:deriv}. These derivatives are a direct measure of the release rate of excess volume and directly comparable to the data obtained from DSC measurements which represent the release rate of excess energy. At temperatures below $535$~K, which is indicated by the dashed vertical line, two annealing processes can be identified. These processes are, as expected, most pronounced at the heating rate of $6$~K/min. The arrows in the figure indicate the positions of the minima. For lower heating rates these minima are less pronounced and shifted to lower temperatures.  

\subsection{Calorimetric study}\label{sec:dsc}

\subsubsection{Isochronal measurements}

\begin{figure}[t]
\includegraphics[width=7cm]{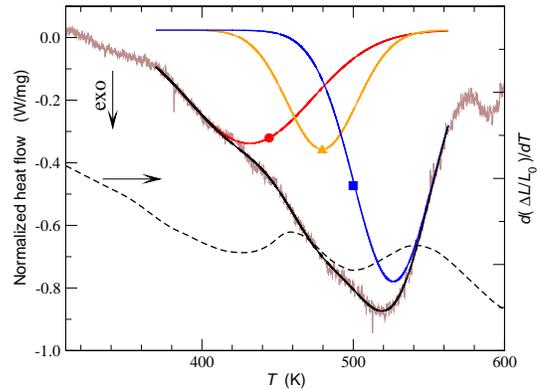}\\
\caption{\footnotesize Data from DSC measurement corresponding to the heating rate of $5$~K/min (brown curve) and its decomposition into three peaks. 
The red (circle), orange (triangle), and blue (square) curves represent fits of three independent Gaussian curves to the DSC data. For comparison, the temperature derivative of the dilatometry curve determined for the heating rate of $6$~K/min is also plotted (dashed line, right ordinate).} \label{fig:DSC}
\end{figure}

\noindent In order to give further insights into the processes contributing to the volume release, DSC measurements were carried out. A typical DSC scan is plotted in Fig.~\ref{fig:DSC} as a function of temperature for the heating rate of $5$~K/min. In the temperature interval from $380$~K to $560$~K, the heat flux can reasonably be presented as a sum of three distinct contributions. The direct comparison with the dilatometry data, of Fig.~\ref{fig:deriv}, most pronounced for the black curve measured with 6K/min at around 420~K, indicates that the first DSC-peak of Fig.~\ref{fig:DSC} is due to annihilation of a distinct type of defect with a release of both, excess free volume and heat. Moreover, a second peak is observed in the dilatometry curve slightly below $500$~K, while two distinct contributions could be recognized for the DSC signal in the temperature range $450~\rm{K} < T < 560$~K.

Accounting for the present results and the literature data \cite{zehet, zehet2} the first peak is attributed to annihilation of single vacancies. The third curve, the blue one, in the DSC signal most likely represents the contribution of dislocation rearrangements and/or their annihilation \cite{zehet2} for a comprehensive discussion of stages III and IV of defect recovery in cold-worked Ni and the effect of impurities see, e.g., Refs. \cite{Seeger, Dlubek}). The second peak in between is most likely related to vacancy--impurity complexes giving release heat does not show significant volumetric changes \cite{Divinski2011, Dlubek}. Comparing the calorimetric and the dilatometry data, it is further obvious that the dislocation reactions, giving rise to a dominant heat release -- the third peak (blue) in the DSC signal around $530$~K -- does not equivalently contribute to a similar strong excess volume release.

The nature of the second, intermediate peak in the DSC signal, which is absent in the dilamometric curve, is not completely clear at present. This signal was not detected in ECAP-deformed pure ($99.99$~wt.\%) Ni \cite{Gerrit} that indicates a contribution of impurity atoms. Moreover, annealing at the corresponding temperature results in a slight increase of the microhardness of ECAP-deformed less pure ($99.6$~wt.\%) Ni, which is investigated in the present paper, that was related to additional dislocation pinning by impurities or vacancy-impurity complexes \cite{Popov}. These facts support the proposed explanation of the intermediate peak on the DSC data. Since it contributes only slightly to the calorimetric signal and is practically invisible in the volumetric data, the corresponding contribution will be neglected in the following analysis.

\subsubsection{Kissinger analysis}

\begin{figure}[ht]
\includegraphics[width=7cm]{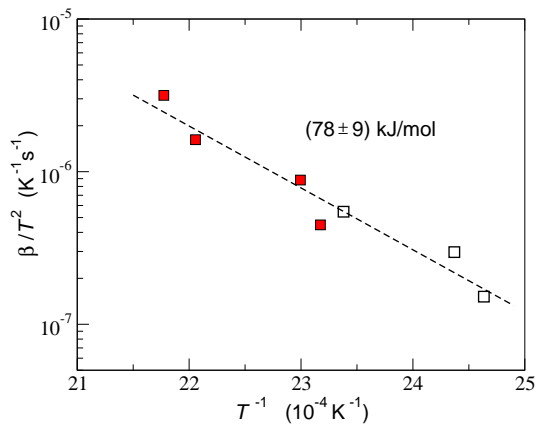}\\
\caption{\footnotesize Kissinger analysis of the DSC (filled squares) and dilatometry (open squares) data for the vacancy peak. The dashed line represents a 'master' fit of the common temperature dependence. The activation enthalpy is indicated.} \label{fig:Kissenger}
\end{figure}

\noindent The calorimetric experiments were performed with different linear heating rates, $\beta$, to investigate the temperature dependence of the heat release and to evaluate the activation enthalpies for the relevant processes. The results of the Kissinger analysis \cite{Kiss} of the first DSC peak, i.e. the plot of $\ln(\beta / T^2_p)$ vs. $T^{-1}_p$, are presented in Fig.~\ref{fig:Kissenger}. Here, $T_p$ is the absolute temperature of the peak at the given heating rate. In Fig.~\ref{fig:Kissenger} the results of Kissinger analysis of the DSC data are plotted together with those obtained from the dilatometry curves. A perfect agreement of the results of the two different experimental methods is seen which allows to draw a single 'master' fit. The resulting activation enthalpy for this process is about $(78 \pm 9)$~kJ/mol that agrees generally with the literature data on the vacancy migration enthalpy in Ni \cite{Wollen, JN}.

The effective activation enthalpies determined from the Kissinger analysis of the second and third DSC peaks are about $(160 \pm 15)$~kJ/mol.

\subsubsection{Isothermal measurements}

\begin{figure*}[t]
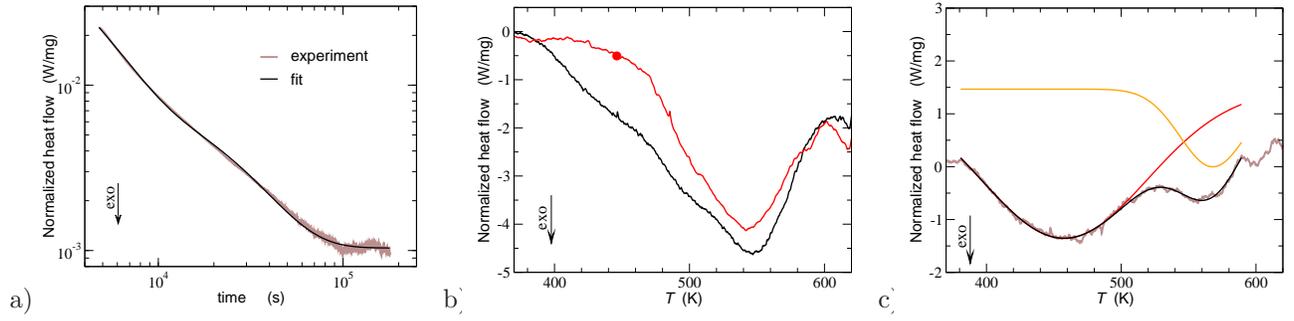

\begin{center}
a) \includegraphics[width=5cm]{figure07a.eps}
\hspace{0.1cm}
b) \includegraphics[width=5cm]{figure07b.eps}
\hspace{0.1cm}
c) \includegraphics[width=5cm]{figure07c.eps} \\
\end{center}
\caption{\footnotesize The heat release (background corrected) during isothermal holding at $390$~K (a). In (b) the DSC curves obtained at a heating rate of $20$~K/min for as-prepared sample (black line) and that after the isothermal holding (red line, circle) are compared and their difference is plotted in (c). The resulting difference signal in (c) could be decomposed into two peaks.
} \label{fig:isotherm}
\end{figure*}

\noindent The nature of the first DSC-peak was further studied using isothermal measurements of the heat release. An ECAP-Ni sample was held at $390$~K for several days and the measured heat release after background subtraction is plotted in Fig.~\ref{fig:isotherm}a. The best fit is given by a sum of two exponential decay functions that correspond to contributions of two processes with relatively short and long time constants. Following the isothermal measurements, the sample was then placed inside the DSC calorimeter and a temperature scan was performed. The two DSC curves measured with the same linear heating rate of $20$~K/min on an as-prepared sample and the specimen undergone the isothermal holding at $390$~K are compared in Fig.~\ref{fig:isotherm}b. As expected, no heat release is seen for the latter sample up to the temperatures slightly above $400$~K. The difference of the two signals is plotted in Fig.~\ref{fig:isotherm}c and it can be decomposed into two peaks. The results are analyzed in detail in the section {\em Defect analysis} below.

\subsection{Effect of pre-annealing on defect volume release}

\noindent In order to determine the excess free volume which was removed during the different stages of the calorimetric study, dilatometry measurements were performed on UFG samples that had been pre-annealed at two selected temperatures.
First, the volumetric signal by heating with a constant rate was determined for a sample that had been pre-annealed at $373$~K for $2$~days (P4). The choice of the annealing temperature was predetermined by the requirement that only one process, i.e. the vacancy annihilation, should contribute to the defect volume release during this annealing treatment and the second process should not have been activated.
Second, one UFG Ni sample was pre-annealed at $480$~K for $17$~hours (P5), i.e., up to that temperature where a substantial reduction of the dislocation density was found in the DSC study.

\begin{figure*}[ht]
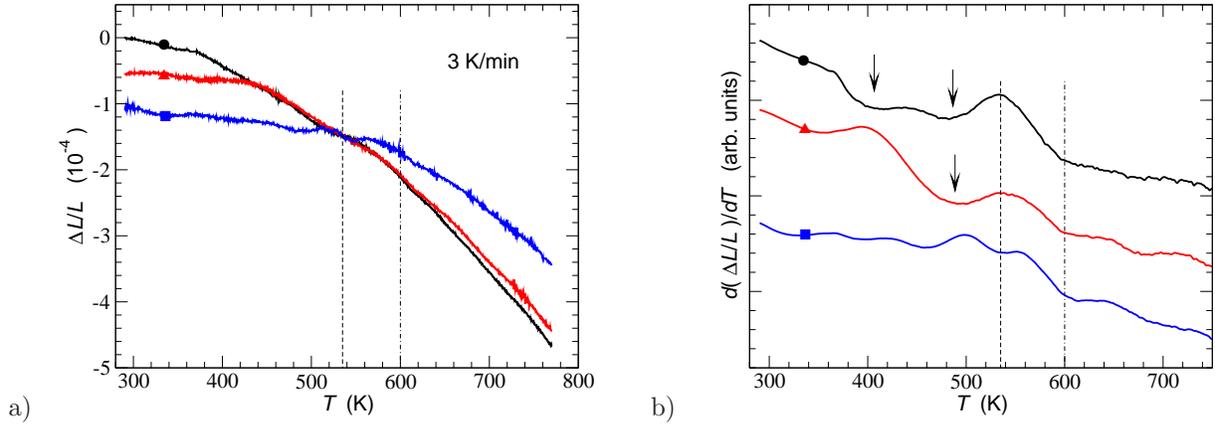

\vspace{1.5cm}

\begin{center}
a)~~ \includegraphics[height=5.5cm]{figure08a.eps}
\hspace{0.5cm}
b)~~~~ \includegraphics[height=5.5cm]{figure08b.eps}\\
\end{center}
\caption{\footnotesize The length changes (a) and their temperature derivatives (b) of an as-prepared (black, circle) and of pre-annealed at $373$~K (P4, red, triangle) and $480$~K (P5, blue, square) ECAP-Ni samples. In (a) the length change curves for the samples P4 and P5 are vertically shifted to match all three signals at $535$~K (see text).} \label{fig:preannealing}
\end{figure*}

The length change curves measured with the heating rate of $3$~K/min on samples P3, P4 and P5 are compared in Fig.~\ref{fig:preannealing}a and the corresponding derivatives, i.e., the transformation rates are given in Fig.~\ref{fig:preannealing}b. The samples pre-annealed at 373~K (P4) and pre-annealed at 480~K (P5) exhibit quite a different length change behavior. Compared to the as-prepared UFG sample without a pre-annealing treatment (black), the total length changes for the samples pre-annealed up to $373$~K are smaller with the sample pre-annealed at 480~K showing the smallest absolute amount of the total released defect volume. Furthermore, a significant length change due to defect annealing occurs at higher temperatures and it starts at about $450$~K and $535$~K for samples pre-annealed at 373~K and at 480~K, respectively. At $T>535$~K (the dashed line in Fig.~\ref{fig:preannealing}) all samples show a qualitatively similar behavior with a two-stage character of the evolution of the temperature derivative $d(\Delta L/L_0)/dT$, Fig.~\ref{fig:preannealing}b. The rate of the length recovery increases rapidly at $535$~K$<T<600$~K, adopting almost a constant value at $T>600$~K. 

For a more detailed analysis, the curves in Fig.~\ref{fig:preannealing} are shifted so that they match (coincide) at the temperature of $535$~K. Note that the shift of dilatometry curves does not affect their derivatives and is a mater of convenience if relative length changes, which correspond to different stages of the excess volume release, are compared for different samples. The choice of $535$~K is the result of the analysis of the derivative curves as shown in Fig.~\ref{fig:deriv} and is more obvious from Fig.~\ref{fig:preannealing}b. Below the temperature of about $535$~K, which is marked by the left vertical dashed line, two different annealing processes can be distinguished in the as-prepared sample (indicated by arrows), see also Fig.~\ref{fig:deriv} for results obtained for various heating rates. For the sample pre-annealed at $373$~K, only one, high-temperature process remains, whereas the sample pre-annealed at $480$~K shows no significant processes related to the volume release below $535$~K. Above $535$~K the behavior of all three samples is qualitatively similar, though an important difference exists: the length changes of as-prepared and annealed at $373$~K samples perfectly coincide above $450$~K, while the defect's volume release occurs at a different rate in the sample pre-annealed at $480$~K, Fig.~\ref{fig:preannealing}a, that is substantiated by a different slope.

The rate of the length change reveals a two-stage character -- one steep stage in the temperature range $535$~K$<T<600$~K and a shallow one with an almost constant value of the derivative $d(\Delta L/L_0)/dT$ for $T>600$~K (Fig.~\ref{fig:preannealing}b). Corresponding kinks in the length changes could be anticipated from Fig.~\ref{fig:preannealing}a, too. The SEM investigation, see below, revealed that the UFG microstructure remains practically stable at temperatures $T<535$~K, whereas grain growth exists at higher temperatures which is affected by the pre-annealing conditions. According to the microstructure analysis, see below, significant grain growth starts at about $535$~K if a specimen is heated with $3$~K/min, although it proceeds differently in samples with different thermal histories. The adopted shifts of the dilatometry curves allowed to quantify the grain growth stage in more detail.

\subsection{Microstructure examination}

\noindent The typical microstructures as observed by EBSD analysis in as-prepared UFG Ni deformed by ECAP at room temperature without pre-annealing (P3) and that of samples pre-annealed at $480$~K for $17$~hours (P5) are shown in Fig.~\ref{fig:GBtrace}a and d, respectively. Only high-angle grain boundaries with misorientation angles $\ge 15^\circ$ are presented. The microstructure of the specimen pre-annealed at 373~K (P4) is unchanged with respect to that of the sample in the as-prepared state without pre-annealing (P3). 

Samples of P3 and P5 were annealed inside the DSC with a heating rate of $3$~K/min up to $600$~K and rapidly cooled down with a rate of $300$~K/min in order to stabilize the microstructures corresponding to the vertical dotted-dashed line in Figs.~\ref{fig:deriv} and \ref{fig:preannealing}b. The microstructures are presented in Figs.~\ref{fig:GBtrace}b and e. Thereafter, the same samples were further annealed inside the DSC first rapidly to $T = 600$~K and then to the final temperature of $770$~K with the linear rate of $3$~K/min, quenched down to room temperature, and then the resulting microstructures, Figs.~\ref{fig:GBtrace}c and f, were analyzed. 

The as-prepared UFG-Ni specimen (P3) is characterized by a plate-like (lamellar) structure aligned along the transverse direction (TD) of ECAP processing in the plane ND--TD, Fig.~\ref{fig:GBtrace}a. The average distance between grain boundaries along the normal direction (ND) is about $(360 \pm 50)$ nm, whereas the grain size in the transverse direction is significantly larger, about several micrometers.

In the perpendicular plane, ND--ED, the high-angle grain boundaries are found to be on average inclined at the angle $\phi \approx (12 \pm 5)^\circ$ to the ECAP direction that is schematically shown in Fig.~\ref{fig:sketch}.

\section{Discussion}

\subsection{Defect analysis}

\noindent 
The total amount of excess volume that is released and that was determined by dilatometry and calorimetry for ECAP-deformed Ni samples with different pre-annealing treatment below 600~K can quantitatively be described by the following equation:    

\begin{equation}
\frac{\Delta V}{V} = C_V \cdot \Omega_V + \gamma_\rho \cdot \rho + e_\gb \cdot S_\gb +  \Delta e_{ne} \cdot S_\gb.
\label{eq:gen}
\end{equation}

\begin{figure*}[ht]
\begin{center}
initial
\hspace{4.cm}
$T=600$~K
\hspace{3.cm}
$T=770$~K \\
\vspace{0.1cm}

a) \includegraphics[width=5cm]{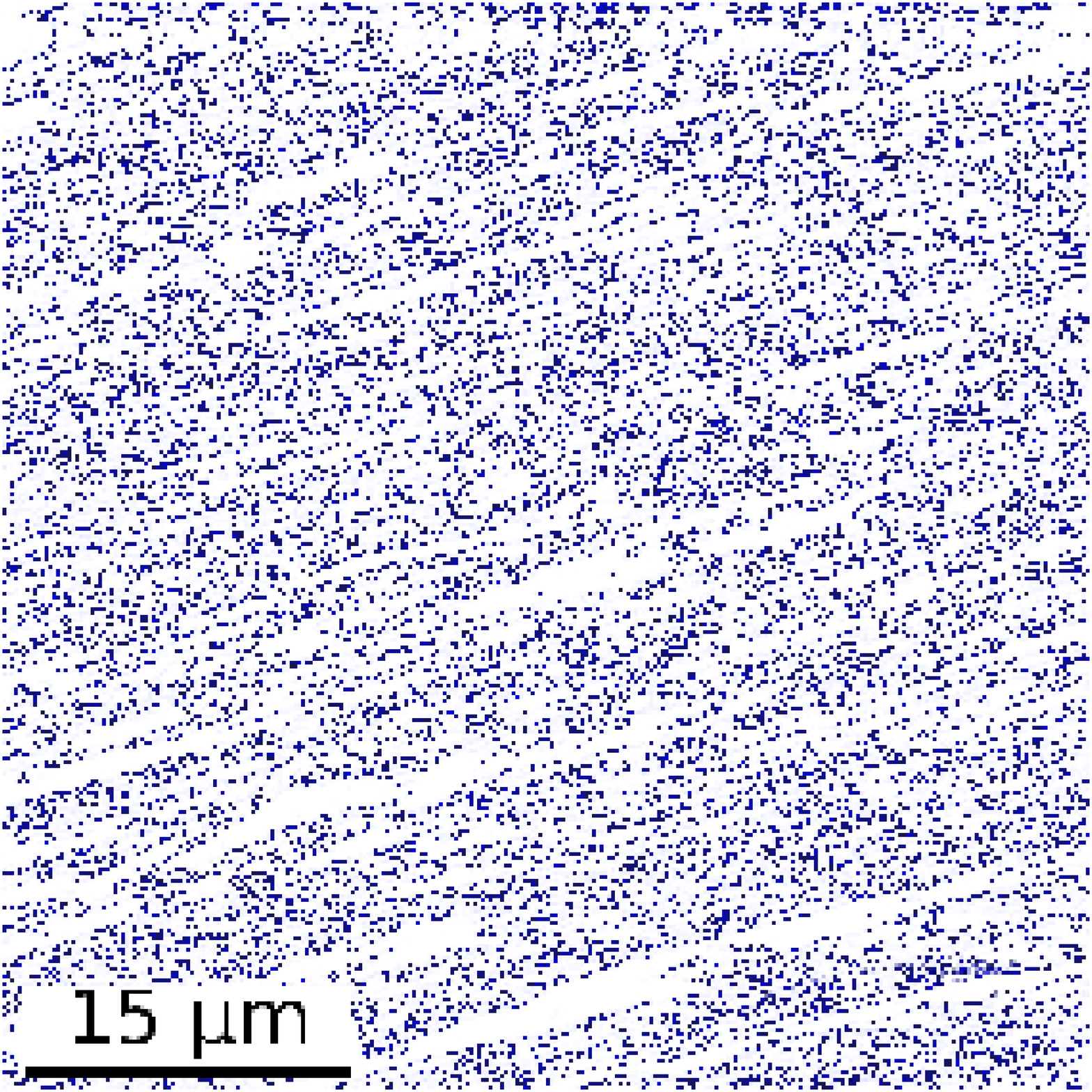}
\hspace{0.1cm}
b) \includegraphics[width=5cm]{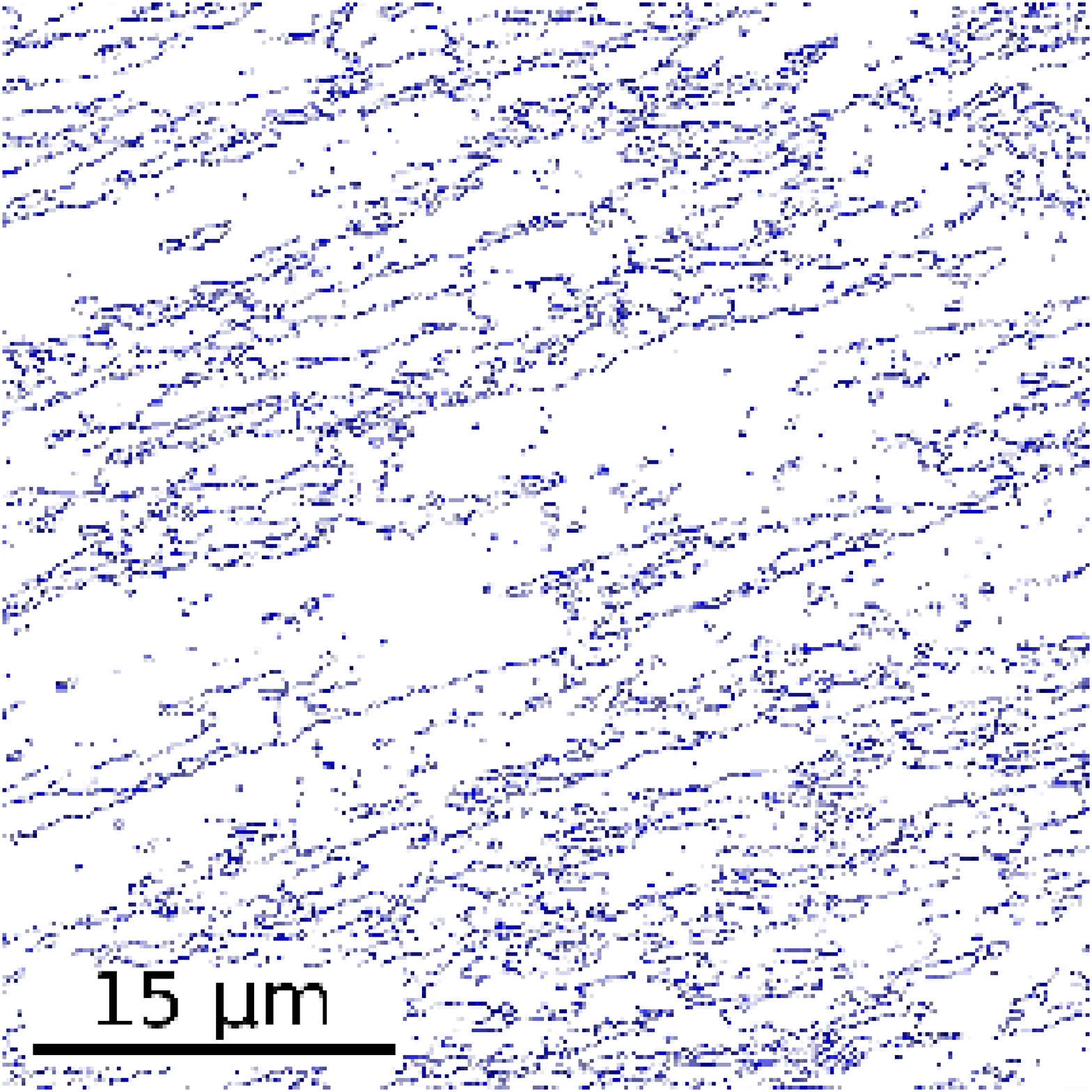}
\hspace{0.1cm}
c) \includegraphics[width=5cm]{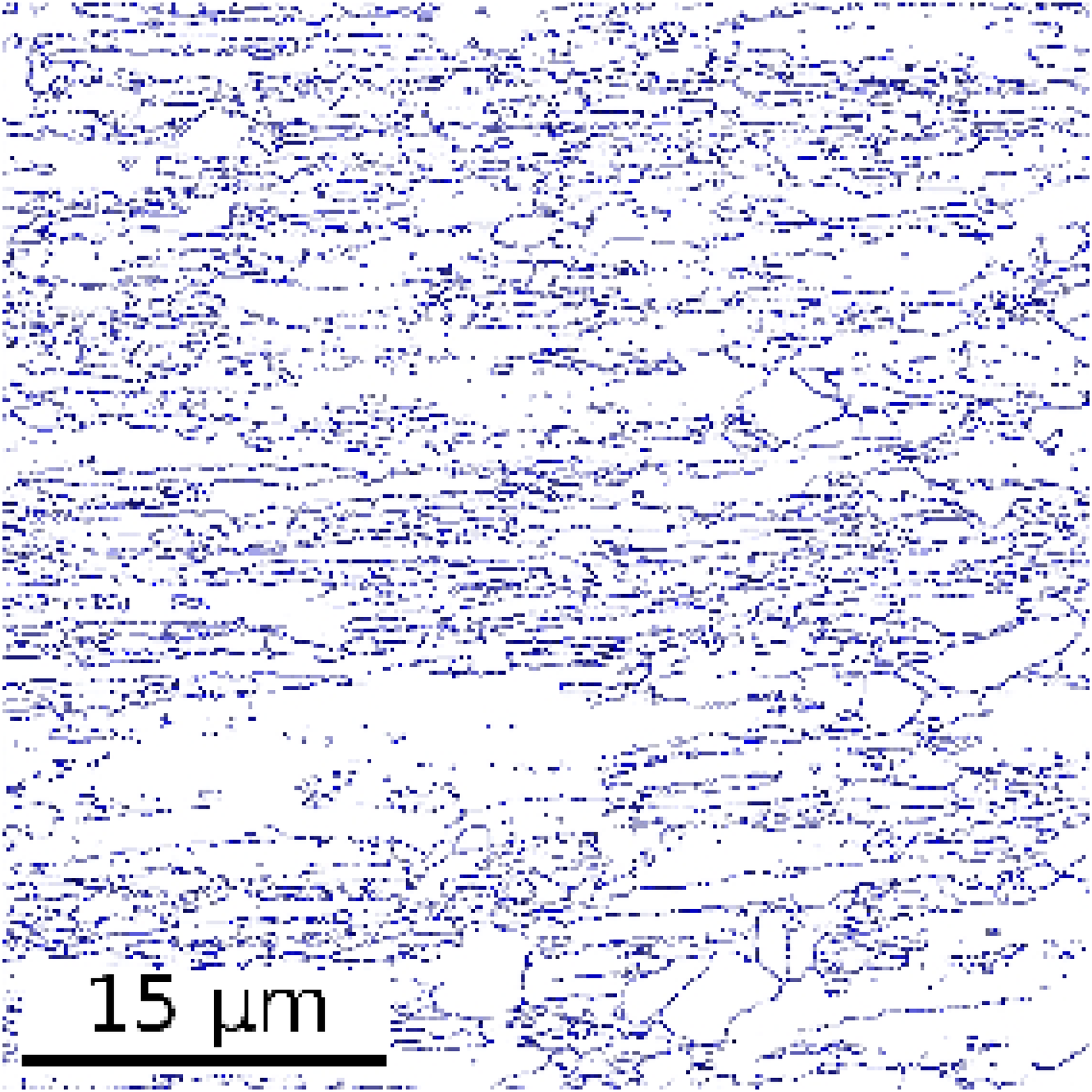} \\

d) \includegraphics[width=5cm]{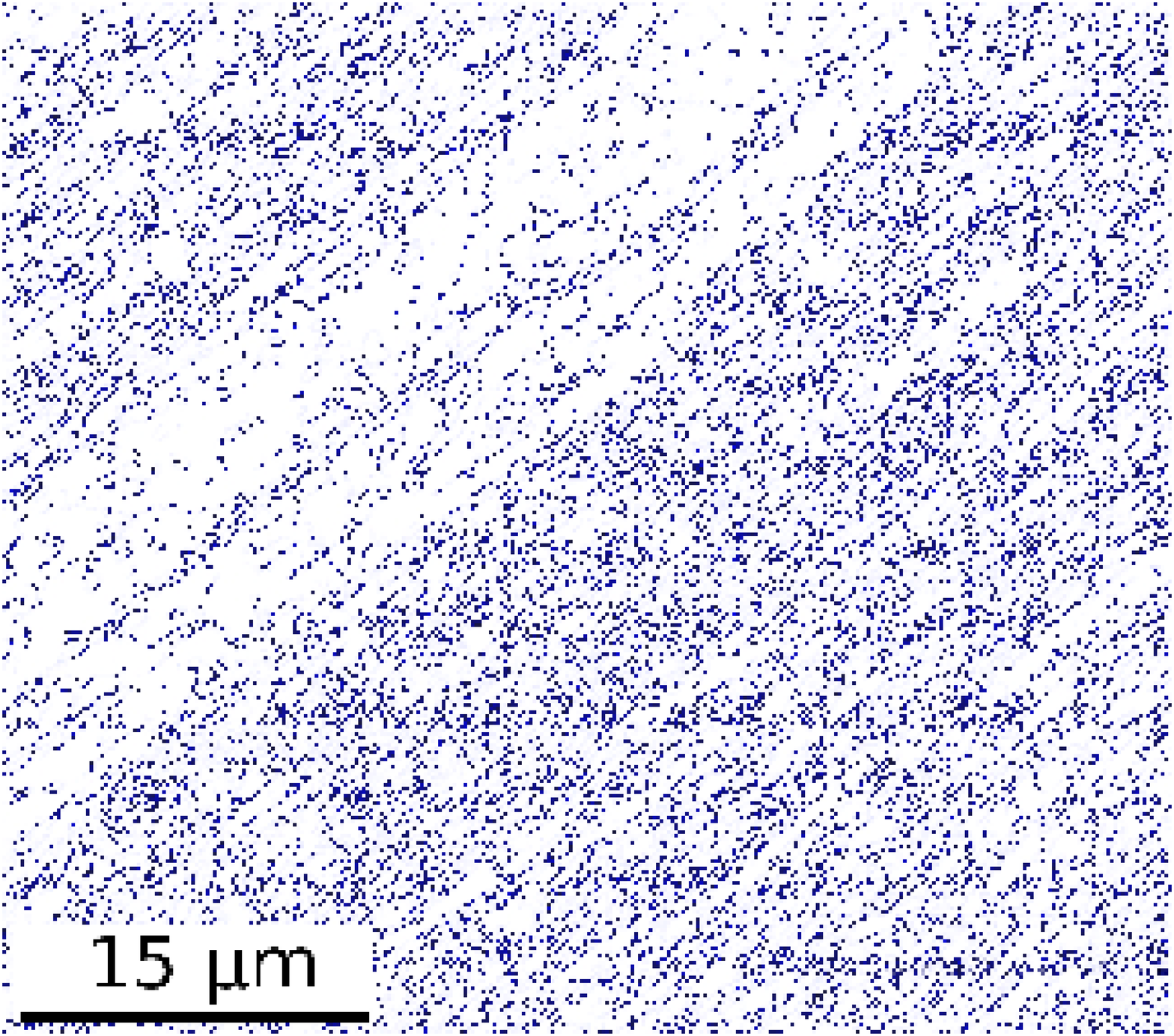}
\hspace{0.1cm}
e) \includegraphics[width=5cm]{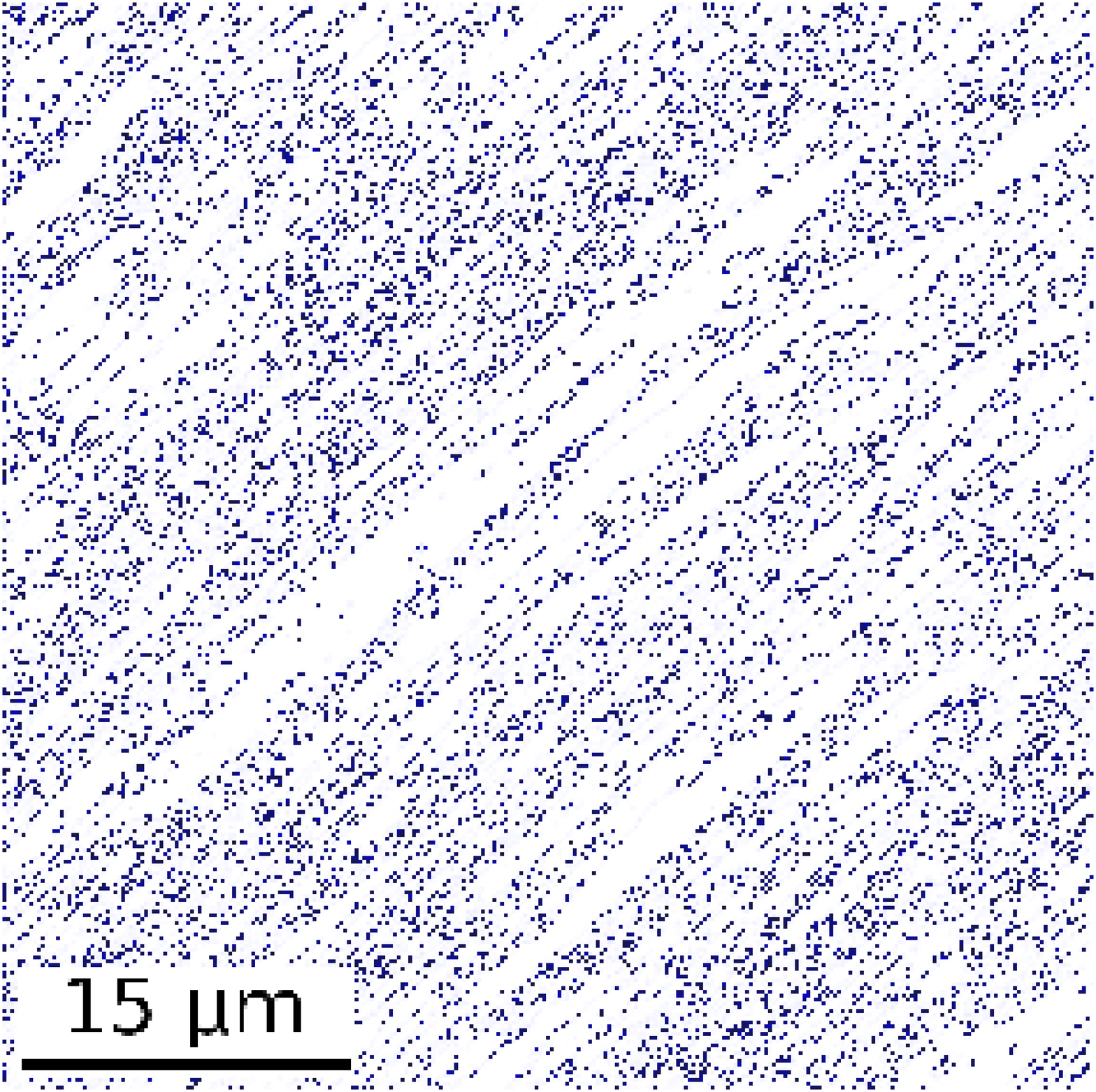}
\hspace{0.1cm}
f) \includegraphics[width=5cm]{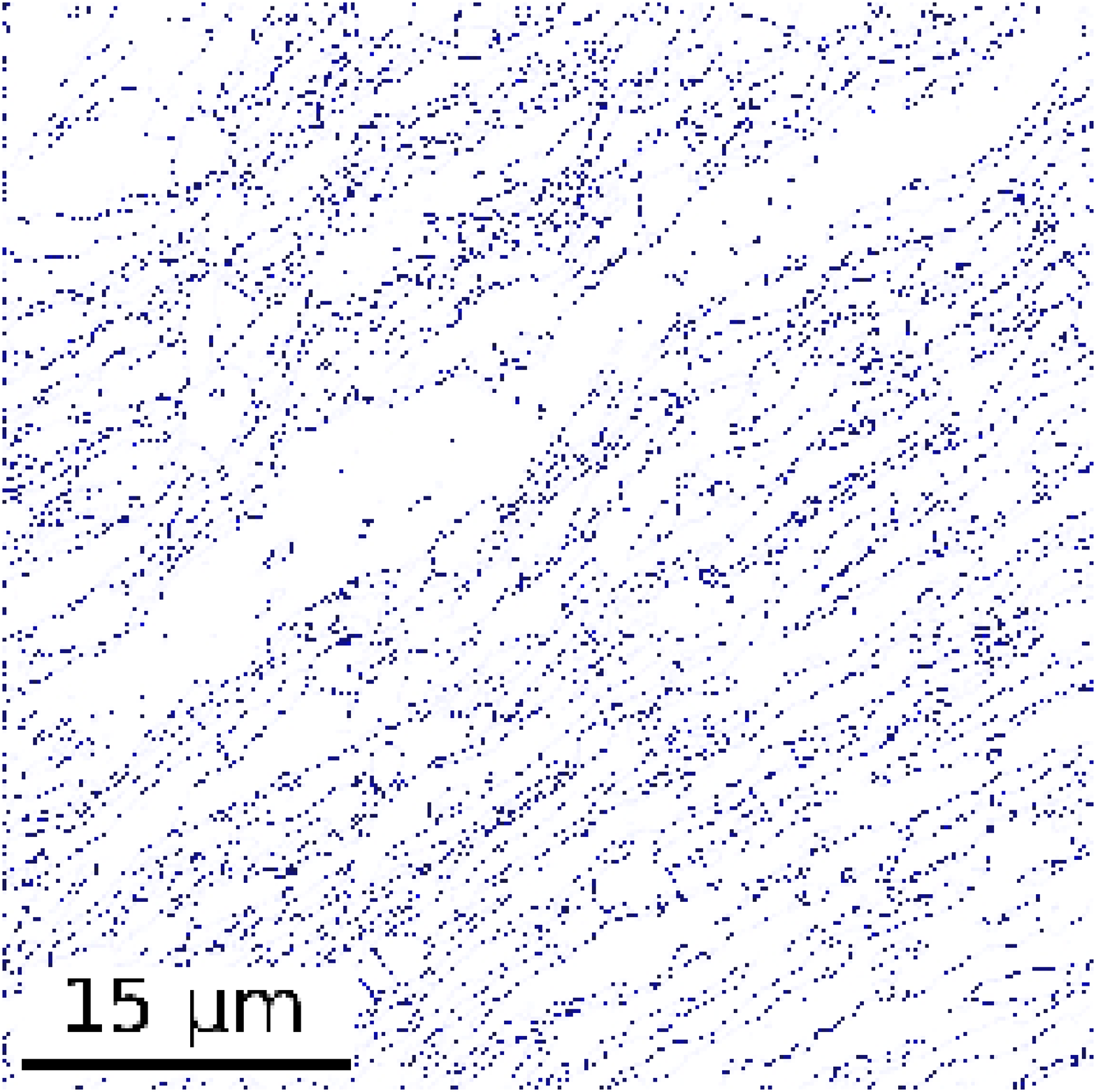}
\end{center}
\caption{\footnotesize High-angle grain boundaries, i.e., boundaries between grains with a misorientation angle $\ge 15^\circ$, determined by orientation imaging microscopy for sample without pre-annealing, P3, (upper row: a, b, c) and the sample pre-annealed at 480~K, P5 (lower row: d, e, f) in the initial state (a, d) and after annealing inside the DSC with a constant heating rate of $3$~K/min to $600$~K (b, e) and to $770$~K (c, f), respectively.} \label{fig:GBtrace}
\end{figure*}

\noindent Here, $C_V$ is the vacancy concentration; $\Omega_V$ is the vacancy formation volume which is a fraction of the atomic volume $\Omega$; $\gamma_\rho$ is the dislocation excess volume per unit length (for an isolated dislocation line, $\gamma_\rho = 0.5 b^2$ with $b$ being the magnitude of the Burgers vector); and $\rho$ is the dislocation density. Furthermore, $S_\gb$ is the volume fraction of the total area of grain boundaries in the material that can be estimated from the analysis of the microstructure 
(see Fig.~\ref{fig:GBtrace}). It has also to be considered that grain boundaries in a UFG material obtained from SPD techniques can be in a 'non-equilibrium' state \cite{ne} and that is characterized by an an increased excess energy, an enhanced diffusivity and an additional free volume, $\left( \delta V/V \right)_{ne} = \Delta e_{ne} \cdot S_\gb$ (see e.g. \cite{Divinski2011}). Here, $\Delta e_{ne}$ is the probable \emph{excess} of grain boundary expansion induced by SPD treatment, such that the net expansion of the 'non-equilibrium' grain boundaries, $e_{ne\gb}$, is defined as $e_{ne\gb} = e_\gb + \Delta e_{ne}$.
Now, in the following it will be shown how the values for $C_V, \rho,  e_\gb $ and $\Delta e_{ne}$ can be derived or at least reasonably be estimated from the experimental data of the present study. 

\subsubsection{Vacancy concentration}

\noindent The combined analysis of volumetric and calorimetric data (Section~\ref{sec:dsc}) suggests that the processes of defect volume release around $400$~K and $500$~K can be associated with annealing of vacancies and dislocations, respectively. The concentration of deformation-induced vacancies, $C_V$, can be derived from the difference of the $\Delta L/L_0$ values between the sample without pre-annealing (P3) and the sample pre-annealed at 373~K (P4), $\left( \frac{\Delta L}{L_0} \right)_{P3-P4} = (0.6 \pm 0.1) \cdot 10^{-4}$, at $293$~K based on the fact that they coincide at $T>480$~K (see Fig.~\ref{fig:preannealing}a). Radiotracer diffusion measurements \cite{Divinski2011} substantiate that the 'non-equilibrium' state of grain boundaries in UFG Ni is not affected by the thermal annealing below $400$~K. Thus, this length change can be solely attributed to the annihilation of deformation-induced vacancies during the pre-annealing of sample P4. However, a simple estimate of the volumetric changes from the variation of the sample length, as it is commonly accepted \cite{Bal, Now, Sprengel2012}, i.e. $\Delta V/V = 3 \Delta L/L$,\footnote{here it is important that $\Delta L$ is a differential signal measured with respect to the reference sample that cancels the contribution of thermal expansion \cite{Sprengel2012}} would not provide data consistent with the independently measured calorimetric effect in terms of the concentration of annihilated vacancies, since this concentration would be significantly overestimated. Therefore, we assume that the vacancies -- isotropic defects by their own -- may produce an anisotropic contribution to the sample shrinkage during their annihilation. This is due to a strong anisotropy of the microstructure, as it was found by the microstructure examination above. 

If a vacancy annihilates at the surface of the sample, an isotropic change of sample's volume is expected. However, in a real polycrystalline specimen vacancies have practically no chance to reach the surface and they annihilate at internal defects, like dislocations and grain boundaries, and each such single defect may provide an \emph{anisotropic} contribution to the volume change. This becomes clear if one imagines a vacancy 'cloud' around a straight dislocation line, then the sample will almost instantaneously shrink perpendicularly to the dislocation line with, in a first approximation, no effect parallel to the dislocation line. Similarly, 'segregated' vacancy layer at a flat grain boundary will result in reduction of the sample size perpendicularly to the grain boundary. It is the random distribution of dislocations and grain boundaries in a typical sample which then 'smear-off' the anisotropic effects from single dislocations and grain boundaries resulting in an isotropic change of sample's volume.

In view of the strongly anisotropic grain shape in ECAP-processed samples, Fig.~\ref{fig:GBtrace}, an anisotropic response on vacancy annihilation may thus be expected. Note that a similar idea was used to explain the diffusion anisotropy of cubic polycrystals which occurred to be induced by an anisotropic distribution of short-circuit paths \cite{dif-ani}.

We assume that a vacancy during its random walk annihilates at a nearest grain boundary. Since the density of grain boundaries along ND is highest, we suggest that these interfaces are dominant sinks for vacancies and an anisotropic length change of the specimen after vacancy annihilation is expected, as it was explained above. (Accounting for low-angle grain boundaries, the average distance between the interfaces in the normal direction will decrease, but it would be still larger than that along the transverse direction.) Since the high-angle grain boundaries are on average inclined at the angle $\phi$ to the ECAP direction in the ND--ED plane,  Fig.~\ref{fig:sketch}, and are parallel to TD in the ND--TD plane, and accounting for the geometry of dilatometric measurements, the following relation is proposed,

\begin{equation}
\left( \frac{\Delta V}{V} \right)_{P3-P4} = C_V \cdot \Omega_V = (\cos \phi + \sin \phi) \left( \frac{\Delta L}{L} \right)_{P3-P4}. 
\label{eq:cv}
\end{equation}

\noindent Here, the indices are related to the sample notation in Table~\ref{table} and $\Omega_V$ is the volume of a relaxed single vacancy. It is accepted that vacancies annihilate at the nearest interfaces which are planes inclined at the angle $\phi$ to the ND--ED plane. In this case, vacancy annihilation contributes to the length changes along ED and ND (the factors $\cos \phi$ and $\sin \phi$ in Eq.~(\ref{eq:cv}), respectively). A contribution along the TD direction is considered here as negligible since the majority of grain boundaries are aligned along this direction, Fig.~\ref{fig:micro}a. Taking a value of $0.82 \Omega$ for the vacancy formation volume in Ni, a vacancy concentration $C_V = (0.9 \pm 0.3) \times 10^{-4}$ is obtained from the dilatometric data. 

From the calorimetric data as shown in Fig~\ref{fig:isotherm} a single, main process running around $400$~K in ECAP-Ni is obvious and is attributed to the heat release due to annealing of vacancies. The measured total heat release corresponding to the first peak is determined to $(0.36 \pm 0.05)$~J/g and taking a value of $1.6$~eV for the vacancy formation enthalpy from the literature \cite{Ni_form_va}, a vacancy concentration of $(1.4 \pm 0.2) \times 10^{-4}$ is deduced. Both values for the vacancy concentration, $C_V$, as determined independently from dilatometry and calorimetry are in good agreement.

\subsubsection{Dislocation density}

\noindent From the DSC data at a temperature slightly higher than the vacancy annealing temperature a heat release attributable to the annealing of dislocations is observed (see blue curve in Fig.~\ref{fig:DSC}). This process is characterized by an amout of about $0.82$~J/g released heat. From this value the forest dislocation density $\rho$ can be derived if the following relation is assumed \cite{zehet, zehet2}

\begin{equation}
 E_{disl} = \frac{Gb^2 \rho}{4 \pi (1-\nu/2)}~ \ln \left( \frac{1}{b \sqrt{\rho}} \right). \label{eq:disl}
\end{equation}

\noindent Here $G$ is the shear modulus, $b$ the Burgers vector, $\nu$ the Poisson ratio, and $E_{disl}$ is the volume density of the stored energy. Using this relation a value for the dislocation density $\rho = 4 \times 10^{15}$~m$^{-2}$ is derived from the calorimetric data. It should be noted that here an equal fraction of edge and screw dislocations was assumed \cite{zehet}.

The estimated dislocation density is typical for SPD-processed metals for which the values from $1$ to $8 \times 10^{15}$~m$^{-2}$ were often reported, see e.g. Refs.~\cite{RV, ES, MZ}.

\subsubsection{Grain boundary expansion and excess of the grain boundary expansion}

\noindent From Fig.~\ref{fig:preannealing} a length change difference of $\left( \Delta L/L \right)_{P4-P5} = (0.6 \pm 0.1) \cdot 10^{-4}$ is obtained for $T< 300$~K for the two samples, pre-annealed at 373~K (P4) and 480~K (P5). This difference is related to substantial recovery of the dislocation density and to relaxation of the deformation-induced state of the grain boundaries in sample P5. The relaxation of grain boundaries is supported by results from diffusion studies \cite{Divinski2011}, where a drastic change of the grain boundary diffusion parameters was observed at a temperature of about $400$~K. Additionally, SEM and EBSD analyses suggest that a fraction of the microstructure, about $15$\%, is recrystallized whereas no grain growth is seen in the other part (see Figs.~\ref{fig:GBtrace}a and d). 

From the dislocation density of about $4\times 10^{15}$~m$^{-2}$ as determined from DSC analysis,
a corresponding value for the release of free volume of $0.9 \times 10^{-4}$ is estimated that translates into a relative length change of $0.4\times 10^{-4}$. Using Eq.~(\ref{eq:gen}), the remaining value, $\left( \Delta L/L \right)_{P4-P5} - \gamma_\rho \cdot \rho \approx 0.2 \times 10^{-4}$, is attributed to relaxation of the deformation-induced 'non-equilibrium' state of high-angle grain boundaries and to an elimination of their certain fraction, $f_{recr} = 0.15$, as a result of recrystallization during pre-annealing. Thus, we obtain the relation

\begin{equation}
0.2 \times 10^{-4} = f_{recr} \cdot \cos \phi \cdot \frac{e_\gb}{d} + \cos \phi \cdot \frac{\Delta e_{ne}}{d}.
\label{eq:P4P5}
\end{equation}

\noindent Here, $d$ is the grain size and it is taken into account that the high-angle grain boundaries are inclined at the angle $90^\circ - \phi$ to the measurement direction. Since the final grain size exceeds several micrometers, its exact value, i.e. the term like $1/d_{end}$, can be neglected due the resolution limits of the dilatometer.

The value of grain boundary expansion $e_\gb$ can be estimated from the volumetric signals measured for samples P3 without and P5 with pre-annealing at 480~K in the range of $535$ and $600$~K, $\left( \Delta L/L \right)^{P3}_{535-600} = (0.6 \pm 0.05) \times 10^{-4}$ and $\left( \Delta L/L \right)^{P5}_{535-600} = (0.25 \pm 0.05) \times 10^{-4}$, since the detailed microstructure data are available, Fig.~\ref{fig:GBtrace}. During linear heating from $535$~K to $600$~K the grain size in, e.g., sample P5 slightly increases from $d_{535} = 360$~nm to $d_{600} = 470$~nm. The corresponding contribution to the linear change of the specimen size is

\begin{equation}
\left( \Delta L/L \right)^{P5}_{535-600} = (1-f_{recr}) \cdot \cos \phi \cdot e_\gb \left( \frac{1}{d_{535}} - \frac{1}{d_{600}} \right).
\label{eq:P5_535_600}
\end{equation}

From Eq. (\ref{eq:P5_535_600}) a value for the grain boundary expansion of $e_\gb = 0.042$~nm is determined. Similar estimates for sample P3 result in the value of $0.037$~nm. Thus, the present experiments substantiate that the grain boundary expansion in Ni is, $e_\gb = (0.039 \pm 0.015)$~nm, as determined independently from previous dilatometric studies on HPT-Ni \cite{Steyskal2012}. The value for $e_\gb$ is attributed to the grain boundary expansion of \emph{relaxed} grain boundaries as they are present in SPD-processed Ni after annealing treatments above $400$~K \cite{Divinski2011} or as can be found in annealed polycrystalline counterpart. 

There is an additional term in Eq.~(\ref{eq:P4P5}) which accounts for the contribution of the 'non-equilibrium' state of high-angle grain boundaries to their excess free volume. Using Eq.~(\ref{eq:P4P5}) we obtain, $\Delta e_{ne} = (0.002 \pm 0.001)$~nm that corresponds to about $7$\% of the relaxed value $e_\gb$. Due to a large uncertainty of this estimation, we conclude that the 'non-equilibrium' state of high-angle grain boundaries is probably characterized by an excess free volume which is hardly measurable and might amount to $5$ to $10$\% of its equilibrium value. There is an additional issue which has to be taken into account. A careful analysis of the grain boundary diffusion data measured on SPD-processed materials, especially in pure Cu and Ni and Cu alloys \cite{R_Act, R_PRL, Amouyal}, resulted in a conclusion that only a fraction of the high-angle grain boundaries, about $10$ to $15$\% is in a 'non-equilibrium' state after ECAP deformation \cite{hierarchy}. Then, the above determined value of $\Delta e_{ne}$ represents, in fact, a lower bond of the extra free volume associated with the 'non-equilibrium' state averaged over all high-angle grain boundaries.

Based on the diffusion data and applying Nazarov's model \cite{Naz, NM} of the 'non-equilibrium' state of grain boundaries, the excess volume release due to relaxation of the 'non-equilibrium' state of grain boundaries was estimated in \cite{Divinski2011} at $0.4\times 10^{-4}$ that is within a factor of two with the present estimates. 

\begin{figure*}[ht]
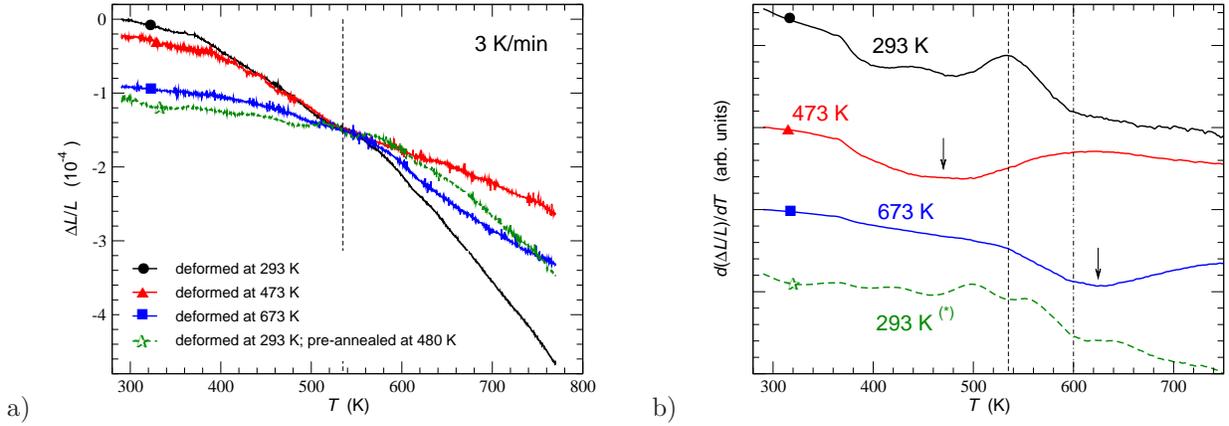

\begin{center}
a)~~ \includegraphics[height=5.5cm]{figure10a.eps}
\hspace{0.5cm}
b)~~~ \includegraphics[height=5.5cm]{figure10b.eps}
\end{center}
\caption{\footnotesize Dilatometric length change curves upon heating with a linear rate of 3~K/min for Ni samples which have undergone ECAP deformation at different temperatures (a) and the corresponding temperature derivatives (b). For comparison the dilatometric curve for the sample that was deformed at room temperature and pre-annealed at $480$~K, P5 (green, star) is shown.} \label{fig:hot}
\end{figure*}

Grain boundary relaxation contributes to the volume release and some contribution to the heat release is to be expected, too. Note that such a contribution could be large, especially in the case of nanocrystals with the grain size of $10$~nm, see e.g. \cite{Tschope}.  In our previous paper, Ref.~\cite{Divinski2011}, we estimated the excess energy related to 'diffusion-modified' state of grain boundaries at about 20 to 30\% of the equilibrium grain boundary energy. Such an effect is perhaps hard to see in the present case of sub-micron-large grains.

\subsubsection{Free volume release at $T > 600$ K}

The present results, furthermore, substantiate a significant free volume release of $\Delta L/L_0 = 2.4 \times 10^{-4}$ between $600$ and $770$~K especially for the samples deformed at ambient temperature. The same amount of the free volume is released in the sample P4 (i.e. deformed at ambient temperature and pre-annealed at $373$) in this temperature interval. A smaller, but still significant amount is released in the sample P5 that was pre-annealed at $480$~K, here the value is about $1.7 \times 10^{-4}$. 

These values cannot be attributed to grain growth. In fact, the average grain size is the same (within the uncertainty limits) in sample P3 at $600$ and $770$~K -- $d^{P3}_{600} = (1170 \pm 300)$~nm and $d^{P3}_{770} = (1200 \pm 200)$~nm, and a distinct grain growth is observed in sample P5, $d^{P5}_{600} = (460 \pm 50)$~nm and $d^{P5}_{770} = (800 \pm 100)$~nm. Using the value of the grain boundary expansion of $e_\gb = 0.039$~nm as determined previously, length changes of less than $1 \times 10^{-5}$ and about $0.3 \times 10^{-4}$ would be expected for samples P3 and P5, for the temperature raise from $600$ to $770$~K. This value is clearly too low to explain the measurements at $T>600$~K just by grain growth.

One possible explanation for the observed contribution to the length change, which was not discussed above, is related to a contribution due to the shrinkage of microvoids at high temperatures. The appearance of the vacancy clustering due to an interaction with impurities was documented for cold-worked Ni for stage III of defect recovery (around $500$~K) \cite{Dlubek-voids} and such a signal is probably registered in the present DSC measurements, Fig.~\ref{fig:DSC}. Dlubek with co-workers found that these microvoids grow first during annealing, but finally annealed out around $700$~K during recrystallization of cold-worked and relatively coarse-grained Ni \cite{Dlubek-voids}.

The existence of individual microvoids in generally dense SPD-processed materials already in as-deformed state was documented, e.g., for ECAP-Ti \cite{Lapovok}. In SPD-processed materials a further defect, open or percolating porosity may exist, that was discovered by special radiotracer experiments for UFG Cu and Cu alloys deformed by ECAP at ambient temperature \cite{R_PRL, R_Scr, R_Act}. Similar measurements found no percolating porosity in the present UFG Ni deformed by ECAP at ambient temperature \cite{Divinski2011}. However, the open porosity appears as a result of annealing treatment of ECAP-Ni \cite{Divinski2011, Divinski2015} or ECAP deformation at elevated temperatures \cite{Gerrit}. One may expect that it is the shrinking of the pores at elevated temperatures which causes additional (irreversible) free volume release of the ECAP-deformed Ni at temperatures $T>600$~K. A model of percolating porosity evolution as a result of grain growth in UFG samples was recently proposed \cite{porosity}. It was shown that under certain conditions the percolating porosity with the volume fraction of several ppms may persist the microstructure recrystallization if initial grain size amounts to several hundreds nanometers or less, whereas the dominant fraction of microvoids sinters during annealing \cite{porosity}. 

Note that there is a significant difference between final microstructures in samples P3 and P5, cf. Figs.~\ref{fig:GBtrace}c) and f), that correlates with a difference of the excess volume release, see Fig.~\ref{fig:preannealing}a). This fact may be related to a  substantial reduction of the defect concentration during the long pre-annealing of sample P5 which hindered grain growth kinetics and perhaps lowered the driving force for recrystallization.

\subsection{Effect of deformation temperature}

\noindent In the following paragraph a qualitative analysis will be given obtained from the dilatometry measurements of speciments which were deformed at elevated temperatures.

Figure~\ref{fig:hot}a shows the dilatometric length change curves $\Delta L / L_0$ upon heating with a linear rate of $3$~K/min for Ni samples that were deformed by ECAP at temperatures of $473$~K and $673$~K. In comparison with the sample deformed at room temperature (293~K, P3), the total release of defect related volume for the hot-deformed samples is significantly smaller and the release sets in at higher temperatures. The total volume release is comparable to that measured on the sample that had been deformed at room temperature and was pre-annealed at $480$~K, Fig.~\ref{fig:hot}a. Note that we have chosen the temperature of $535$~K as an onset of recrystallization and shifted all curves to level up at this temperature.

The maximum rate of free volume release in ECAP-Ni deformed at $473$ (red) and $673$~K (blue) are observed at about $460$ and $630$~K, respectively (indicated by arrows in Fig.~\ref{fig:hot}b). On the other hand, for the sample deformed at room temperature and annealed at $480$~K (green) the amount of free volume release is most  pronounced above $535$~K, whereas the total volume released between $300$ and $770$~K is almost of the same amount as for the three samples, Fig.~\ref{fig:hot}a. It suggests that the warm deformation at $473$~K introduces dislocations which are removed by annealing at almost the same temperature. Deformation at an even higher temperature, $673$~K, results in dislocation configurations for which a higher thermal activation for removal is required and with a small effect of grain growth. On the other hand, grain growth during linear heating is less pronounced in the samples deformed at 473~K (P6) and at 673~K (P7) (the average grain size amounts to about $2$ to $5~\mu$m even in the initial, post-deformation state), while grain growth is significant in the sample deformed at room temperature and annealed at 480~K (P5) before the dilatometry measurements. 

The defect volume release in the sample deformed at 473~K (P6) starts at a temperature about $30$~K higher with respect to the sample that was deformed at 293~K (P3) and then the two dilatometry curves run almost parallel up to the temperature of $500$~K, Fig.~\ref{fig:hot}a. This fact indicates a similar amount of dislocation-type defects released in both samples in this temperature interval and a smaller concentration of stored vacancies in the sample deformed at 473~K (P6). The difference in the length changes at $380$~K between samples P3 and P6 (i.e. the black and red curves in Fig.~\ref{fig:hot}a, respectively) is about $0.2 \times 10^{-4}$ and it  translates into the difference between the corresponding concentration of deformation-introduced vacancies of about $2.4 \times 10^{-5}$. A vacancy-related peak could hardly be revealed for the length changes of the sample P7 (i.e. the blue line in Fig.~\ref{fig:hot}a) and is below the resolution limit of the present experiments.

\section{Summary and Conclusions}

\noindent In the present investigation it is demonstrated for ECAP-processed Ni that volumetric and stored energy studies as realized by difference dilatometry and differential scanning calorimetry in combination with detailed microstructural characterization are complementary approaches ideally suited for the in-depth analysis of atomic defects. Besides the results given in the preceding section the following main aspects are to be highlighted.

\begin{itemize}
 \item The concentration of deformation-induced vacancies in ECAP-Ni is about $10^{-4}$ when deformation is performed at room temperature via the route $\BC$ and it is decreased with increasing temperature of deformation.

 \item The grain boundary expansion in Ni equals to $(0.039 \pm 0.015)$~nm.

 \item SPD deformation increases the excess volume related to grain boundaries. The absolute value of such an excess of the grain boundary expansion is small and does not exceed 5 to 10\%.
\end{itemize}

The study has been performed on technically pure Ni that allowed qualitative separation of various stages of defect evolution which can be prescribed by specific deformation-induced defects. In the previous research on grain boundary diffusion in the same material \cite{Divinski2011}, Si has been identified as a critic impurity which affects interface thermodynamics, although the role of other impurities, Table~\ref{tab:impurity}, may be important, too. A similar study on specific alloying systems would be helpful and this is a subject of on-going research.
  
\subsection*{Acknowledgment}

The authors are grateful to Prof. Y. Estrin and Dr. R. Lapovok (Monash University, Clayton, Australia) for fruitful collaboration and support in ECAP experiments. Assistance by J. Bokeloh (Institute of Materials Physics, University of M\"unster) with the DSC measurements is appreciated. The financial support by DFG (Deutsche Forschungsgemeinschaft) project WI 1899/9-2 and by the Austrian Science Fund (FWF) projects P21009-N20 and P25628-N20 are acknowledged.


\begin{thebibliography}{100}

\singlespacing
\bibitem{book}R.Z. Valiev, M.J. Zehetbauer, Y. Estrin, H.W. H\"oppel, Y. Ivanisenko, H. Hahn, G. Wilde, H.J. Roven, X. Sauvage, T.G. Langdon, The innovation potential of bulk nanostructured materials, Adv Engn Mater 9 (2007) 527-533.

\bibitem{Divinski2011}S.V. Divinski, G. Reglitz, H. R\"osner, Y. Estrin, G. Wilde, Self-diffusion in Ni prepared by severe plastic deformation: effect of non-equilibrium grain boundary state, Acta Mater 59 (2011) 1974-1985.

\bibitem{Divinski2015}S.V. Divinski, G. Reglitz, I. Golovin, M. Peterlechner, G. Wilde, Effect of heat treatment on diffusion, internal friction, microstructure and mechanical properties of ultrafine grained nickel severely deformed by equal channel angular pressing , Acta Mater 82 (2015) 11-21.

\bibitem{Divinski2014}S.V. Divinski, G. Reglitz, M. Wegner, M. Peterlechner, G. Wilde, Effect of pinning by orientation gradient on the thermal stability of ultrafine grained Ni produced by Equal Channel Angular Pressing, J Appl Phys 115 (2014) 113503.

\bibitem{zehet}D. Setman, E. Schafler, E. Korznikova, M.J. Zehetbauer, The presence and nature of vacancy type defects in nanometals detained by severe plastic deformation, Mater Sci Eng A 493 (2008) 116-122.

\bibitem{zehet2}D. Setman, M.B. Kerber, E. Schafler, M.J. Zehetbauer, Activation enthalpies of deformation-induced lattice defects in severe plastic deformation nanometals measured by differential scanning calorimetry, Metall Mater Trans A 41 (2010) 810-815.

\bibitem{Sprengel2012}W. Sprengel, B. Oberdorfer, E.M. Steyskal, R. W{\"u}rschum, Dilatometry: a powerful tool for the study of defects in ultrafine-grained metals, J Mater Sci 47 (2012) 7921-7925.

\bibitem{Steyskal2012}E.M. Steyskal, W. Sprengel, B. Oberdorfer, M.J. Zehetbauer, R. Pippan, R. W\"urschum, Direct Experimental Determination of Grain Boundary Excess Volume in Metals, Phys Rev Lett 108 (2012) 055504.

\bibitem{OberdorferPRL}B. Oberdorfer, E.M. Steyskal, W. Sprengel, W. Puff, P. Pikart, C. Hugenschmidt, M.J. Zehetbauer, R. Pippan, R. W{\"u}rschum, In Situ Probing of Fast Defect Annealing in Cu and Ni with a High-Intensity Positron Beam, Phys Rev Lett 105 (2010) 146101.

\bibitem{Oberdorfer2011}B. Oberdorfer, E.M. Steyskal, W. Sprengel, R. Pippan, M. Zehetbauer, W. Puff, R. W{\"u}rschum, Recrystallization kinetics of ultrafine-grained Ni studied by dilatometry, J Alloys Compd 509 (2011) S309-S311. 

\bibitem{Cu} B. Oberdorfer, D. Setman, E.M. Steyskal, A. Hohenwarter, W. Sprengel, M.J. Zehetbauer, R. Pippan, R. W\"urschum, Grain boundary excess volume and defect annealing of copper after high-pressure torsion, Acta Mater 68 (2014) 189-195.

\bibitem{Kotzurek2013}J. Kotzurek, Master Thesis, Graz University of Technology, Austria (2013).

\bibitem{Oberdorfer2010}B. Oberdorfer, B. Lorenzoni, K. Unger, W. Sprengel, M. Zehetbauer, R. Pippan, R. W{\"u}rschum, Absolute concentration of free volume-type defects in ultrafine-grained Fe prepared by high-pressure torsion, Scr Mater 63 (2010) 452-455.

\bibitem{Pippan1}E. Schafler, R. Pippan, Effect of thermal treatment on microstructure in high pressure torsion (HPT) deformed nickel, Mater Sci Eng A 387-389 (2004) 799-804.

\bibitem{Pippan2}H.W. Zhang, X. Huang, R. Pippan, N. Hansen, Thermal behavior of Ni (99.967\% and 99.5\% purity) deformed to an ultra-high strain by high pressure torsion, Acta Mater 58 (2010) 1698-1707.

\bibitem{Humphreys}F.J. Humphreys, M. Hatherly, Recrystallization and Related Annealing Phenomena (2nd Ed.), Oxford, Elsevier (2004). 

\bibitem{Seeger}A. Seeger, D. Schumacher, W. Schilling, J. Diehl, (Ed.) Vacancies and Interstitials in Metals, North-Holland Publ. Co., Amsterdam (1970).

\bibitem{Dlubek}G. Dlubek, O. Br\"ummer, P. Sickert, Study of the nature of the processes occurring in the recovery  stages iii and iv of plastically deformed ni of different purity by means of the positron annihilation method, phys. stat. sol. (a) 89, 401-410 (1977) 

\bibitem{Gerrit}G. Reglitz, PhD Thesis, M{\"u}nster University, Germany (2013).

\bibitem{Popov}V.V. Popov, E.N. Popova, D.D. Kuznetsov, A.V. Stolbovsky, E.V. Shorohov, G. Reglitz, S.V. Divinski, G. Wilde, Evolution of Ni structure at dynamic channel-angular pressing, Mater Sci Eng A 585 (2013) 281-291.

\bibitem{Kiss}H. Kissinger, Reaction kinetics in differential thermal analysis, Analyt Chem 29 (1957) 1702-1706.

\bibitem{Wollen}H. Wollenberger, Point Defects, in: Cahn R, Haasen P. (Eds.). Physical Metallurgy, Amsterdam: North Holland (1986).

\bibitem{JN}W. Wycisk, M. Feller-Kniepmeier, Quenching experiments in high-purity Ni, J Nucl Mater 69-70 (1978) 616-619.

\bibitem{ne}X. Sauvage, G. Wilde, S.V. Divinski, Z. Horita, R.Z. Valiev, Grain boundaries in ultrafine grained materials processed by severe plastic deformation and related phenomena, Mater Sci Eng A 540 (2012) 1-12.

\bibitem{Bal}R. Simmons, R. Balluffi, Measurements of equilibrium vacancy concentrations in aluminum, Phys Rev 117 (1960) 52-61.

\bibitem{Now}R. Feder, A. Nowick, Use of thermal expansion measurements to detect lattice vacancies near the melting point of pure lead and aluminum, Phys Rev 109 (1958) 1959-1963.

\bibitem{dif-ani}S.V. Divinski, L.N. Larikov, On the Diffusion Anisotropy of Polycrystals, Cryst Reserch Technol 30 (1995) 957-962.

\bibitem{Ni_form_va}K. Ehrhart, in: Landolt-B{\"o}rnstein, New Series III/25. Atomic Defects in Metals (1991), p. 243.

\bibitem{RV}R.Z. Valiev, I.V. Alexandrov, Y.T. Zhu, T. Lowe, Paradox of strength and ductility in metals processed by severe plastic deformation, J Mater Research 17 (2002) 5-8.

\bibitem{ES}E. Schafler, G. Steiner, E. Korznikova, M. Kerber, M.J. Zehetbauer, Lattice defect investigation of ECAP-Cu by means of X-ray line profile analysis, calorimetry and electrical resistometry, Mater Sci Engineer A 410–411 (2005) 169–173.

\bibitem{MZ}M.J. Zehetbauer, H.P. St\"uwe, A. Vorhauer, E. Schafler, J. Kohout, The Role of Hydrostatic Pressure in Severe Plastic Deformation,  Adv Engineer Mater 5 (2003) 330-337.

\bibitem{R_Act}S.V. Divinski, J. Ribbe, D. Baither, G. Schmitz, G. Reglitz, H. Rösner, K. Sato, Y. Estrin, G. Wilde, Nano- and micro-scale free volume in ultrafine grained Cu--1wt\%Pb alloy deformed by equal channel angular pressing, Acta Mater 57 (2009) 5706-5717.

\bibitem{R_PRL}J. Ribbe, D. Baither, G. Schmitz, S.V. Divinski, Cracks in ultra fine grain copper produced by equal channel angular pressing, Phys Rev Lett 102 (2009) 165501.

\bibitem{Amouyal}Y. Amouyal, S.V. Divinski, Y. Estrin, E. Rabkin, Short-circuit diffusion in an ultrafine grain copper-zirconium alloy produced by equal channel angular pressing, Acta Mater 55 (2007) 5968-5979.

\bibitem{hierarchy}S.V. Divinski, G Wilde, Diffusion in ultra-fine grained materials, Mater Sci Forum 584-586 (2008) 1012-1018.

\bibitem{Naz}A.A. Nazarov, A.E. Romanov, R.Z. Valiev, On the structure, stress-fields and energy of nonequilibrium grain-boundaries, Acta Metall Mater 41 (1993) 1033-1040.

\bibitem{NM}A.A. Nazarov, R.R. Mulyukov, Nanostructured materials, in: Goddard III WA, Brenner DW, Lyshevski SE, Iafrate GJ (Eds.). Handbook of Nanoscience, Engineering, and Technology, Boca Raton: CRC Press (2002).

\bibitem{Tschope} A. Tsch\"ope, R. Birringer, H. Gleiter, Calorimetric measurements of the thermal relaxation in nanocrystalline platinum, J Appl Phys 71 (1992) 5391-5394. 

\bibitem{Dlubek-voids}G. Dlubek, O. Br\"ummer, N. Meyendorf, P. Hautoj\"arvi, A. Vehanen, J. Yli-Kauppila, Impurity-induced  vacancy clustering in cold-worked nickel, J Phys F: Metal  Phys 9 (1979) 1961-1973.

\bibitem{Lapovok}R. Lapovok, D. Tomus, J. Mang, Y. Estrin, T.C. Lowe, Evolution of nanoscale porosity during equal-channel angular pressing of titanium,  Acta Mater 57 (2009) 2909-2918.

\bibitem{R_Scr}J. Ribbe, D. Baither, G. Schmitz, S.V. Divinski, Ultra fast diffusion and internal porosity in ultra fine grain copper–lead alloy prepared by equal channel angular pressing, Scr Mater 61 (2009) 129-132.

\bibitem{porosity}J. Ribbe, G. Schmitz, D. Gunderov, Y. Estrin, Y. Amouyal, G. Wilde, S.V. Divinski, Effect of annealing on percolating porosity developed in ultrafine grained copper produced by Equal Channel Angular Pressing, Acta Mater 61 (2013) 5477-5486.

\end{thebibliography}
\end{document}